\documentstyle[preprint,eqsecnum,aps]{revtex}
\begin{document}
\title{On the Zero-Point Energy of a Conducting Spherical Shell} 
\author{Giampiero Esposito,$^{1,2}$
\thanks{Electronic address: giampiero.esposito@na.infn.it}
Alexander Yu. Kamenshchik$^{3,4}$
\thanks{Electronic address: kamen@landau.ac.ru}
and Klaus Kirsten$^{5}$
\thanks{Electronic address: klaus.kirsten@itp.uni-leipzig.de}}
\address{${ }^{1}$Istituto Nazionale di 
Fisica Nucleare, Sezione di Napoli,\\
Mostra d'Oltremare Padiglione 20, 80125 Napoli, Italy\\
${ }^{2}$Dipartimento di Scienze Fisiche,\\
Mostra d'Oltremare Padiglione 19, 80125 Napoli, Italy\\
${ }^{3}$L. D. Landau Institute for Theoretical
Physics of Russian Academy of Sciences,\\
Kosygina Str. 2, Moscow 117334, Russia\\
${ }^{4}$Landau Network-Centro Volta, Villa Olmo,\\
Via Cantoni 1, 22100 Como, Italy\\
${ }^{5}$Universit\"{a}t Leipzig, 
Institut f\"{u}r Theoretische Physik,\\
Augustusplatz 10, 04109 Leipzig, Germany}
\maketitle
\begin{abstract}
The zero-point energy of a conducting spherical shell is
evaluated by imposing boundary conditions on the potential
$A_{\mu}$, and on the ghost fields. The scheme requires
that temporal and tangential components of $A_{\mu}$ perturbations
should vanish at the boundary, jointly with the gauge-averaging
functional, first chosen of the Lorenz type. Gauge invariance
of such boundary conditions is then obtained provided that the
ghost fields vanish at the boundary. Normal and longitudinal
modes of the potential
obey an entangled system of eigenvalue equations,
whose solution is a linear combination of Bessel functions
under the above assumptions, and with the help of the Feynman
choice for a dimensionless gauge parameter. 
Interestingly, ghost modes cancel exactly the contribution to
the Casimir energy resulting from transverse and temporal
modes of $A_{\mu}$, jointly with the decoupled 
normal mode of $A_{\mu}$.
Moreover, normal and longitudinal components of $A_{\mu}$ for the
interior and the exterior problem give a result in complete
agreement with the one first found by Boyer, who studied 
instead boundary conditions involving TE and TM modes of the
electromagnetic field. The coupled eigenvalue equations for
perturbative modes of the potential are also analyzed in the
axial gauge, and for arbitrary values of the gauge parameter.
The set of modes which contribute to the Casimir energy is then
drastically changed, and comparison with the case of a flat 
boundary sheds some light on the key features of the Casimir
energy in non-covariant gauges. 

\end{abstract}
\section*{1. Introduction}
\setcounter{section}{1}
One of the most striking applications of quantum field theory
to physical processes consists of the analysis of zero-point
energies.$^{1-7}$ 
Indeed, it is by now well known that zero-point
energies of quantized fields are infinite, 
whereas suitable differences
in zero-point energies are finite and measurable. A good example
is provided by the quantum theory of the electromagnetic field.
In that case, the energy of two perfectly conducting plates
of area $A$ and separation $d$ is found to be$^1$ 
\begin{equation}
\bigtriangleup E=-{\pi^{2}{\hbar}c A \over 720 d^{3}} .
\end{equation}
One thus has an attraction between such plates. 
Further to this effect,
the zero-point energy of a conducting spherical shell was
calculated in Ref. 2 for the first time, giving, however, a
different sign with respect to the case of parallel plates:
\begin{equation}
\bigtriangleup E (R) \cong 0.09 {{\hbar}c 
\over 2R } ,
\end{equation}
where $R$ 
is the radius of the sphere. 
The result (1.2) has been confirmed, independently, by all
authors working in this research field. A powerful
tool turns out to be the method relying on the use of the
generalized $\zeta$-function. The boundary
conditions of the problem (see below) are of the Dirichlet
or Robin type. The corresponding Laplace-type operators, 
say $Q$, are then self-adjoint, and the analytic continuation
of their $\zeta$-function:
\begin{equation}
\zeta_{Q}(s) \equiv {\text {Tr}}_{L^{2}} (Q^{-s})
\end{equation}
may be used to obtain the value of 
$\bigtriangleup E(R)$.$^{6,8}$
We here omit all details concerning 
$\zeta$-function calculations, and we focus instead on the
problem of boundary conditions. 

Let $r,\theta, \varphi$ be the spherical coordinates, which
are appropriate for the analysis of a conducting spherical
shell. The boundary conditions require that the tangential
components of the electric field should vanish, i.e.,
\begin{equation}
[E_{\theta}]_{\partial M}=0 ,
\end{equation}
\begin{equation}
[E_{\varphi}]_{\partial M}=0 .
\end{equation}
These boundary conditions are gauge-invariant, since they 
involve some components of the electromagnetic-field tensor:
\begin{equation}
F_{\mu \nu} \equiv \nabla_{\mu}A_{\nu}
-\nabla_{\nu}A_{\mu} ,
\end{equation}
which is gauge-invariant whenever the connection $\nabla$
is torsion-free, as we always assume. The modes of the
electromagnetic field are then split into transverse electric
(TE) or magnetic multipole:$^9$
\begin{equation}
{\vec r} {\cdot} {\vec E}=E_{r}=0 ,
\end{equation}
and transverse magnetic (TM) or electric multipole:$^9$ 
\begin{equation}
{\vec r} {\cdot} {\vec B}=B_{r}=0 .
\end{equation}
On using the notation in Ref. 9 for spherical Bessel
functions, and the formulae therein for $E_{\theta}$ and
$E_{\varphi}$, one finds that, in the TE case, Eqs. (1.4)
and (1.5) lead to
\begin{equation}
\Bigr[A_{l} j_{l}(kr)+B_{l} n_{l}(kr)\Bigr]_{\partial M}=0,
\end{equation}
while in the TM case the vanishing of $E_{\theta}$ and 
$E_{\varphi}$ at the boundary implies that
\begin{equation}
\left[{d\over dr}(r(C_{l} j_{l}(kr)+D_{l} n_{l}(kr)))
\right]_{\partial M}=0 .
\end{equation}
Of course, if the background includes the point $r=0$, the
coefficients $B_{l}$ and $D_{l}$ should be set to zero for
all values of $l$ to obtain a regular solution. Such a 
singularity is instead avoided if one studies the annular
region in between two concentric spheres.

The boundary conditions in the form (1.9) and (1.10) are
sufficient to implement the algorithm of the $\zeta$-function.
However, they do not tell us anything about the contribution
of the ghost field. Indeed, if one follows a path-integral
approach to the quantization of Maxwell theory, it is 
well-known that the second-order operator acting on
$A^{\mu}$ perturbations when the Lorenz gauge-averaging 
functional is chosen turns out to be, in a flat background,$^7$ 
\begin{equation}
P_{\mu \nu}=-g_{\mu \nu} \Box 
+\left(1-{1\over \alpha} \right) \nabla_{\mu} \nabla_{\nu} .
\end{equation}
With a standard notation, $g$ is the background metric, 
$\alpha$ is a dimensionless parameter,
$\Box$ is the D'Alembert operator
$$
\Box=-{\partial^{2}\over \partial t^{2}}
+ \bigtriangleup ,
$$
where $\bigtriangleup$ is the Laplace operator 
(in our problem, $\bigtriangleup$ is considered on a disk). 
This split of the $\Box$ operator leads eventually to the
eigenvalue equations for the Laplace operator acting on the
temporal, normal and tangential components of the potential, 
after using the equations (2.1)--(2.3) of Sec. 2.
At this stage, the potential, with its gauge transformations
\begin{equation}
{ }^{\varepsilon}A_{\mu} \equiv A_{\mu}+
\nabla_{\mu} \varepsilon ,
\end{equation}
is viewed as a more fundamental object. Some care,
however, is then necessary to ensure gauge invariance of the
whole set of boundary conditions. For example, if one imposes
the boundary conditions
\begin{equation}
[A_{t}]_{\partial M}=0 ,
\end{equation}
\begin{equation}
[A_{\theta}]_{\partial M}=0 ,
\end{equation}
\begin{equation}
[A_{\varphi}]_{\partial M}=0 ,
\end{equation}
this is enough to ensure that the boundary conditions (1.4)
and (1.5) hold. However, within
this framework, one has to impose yet another  
condition. By virtue of Eq. (1.12), the desired boundary
condition involves the gauge function:
\begin{equation}
[\varepsilon ]_{\partial M}=0 .
\end{equation}
Equation (1.16) ensures that the boundary conditions
(1.13)--(1.15) are preserved under the gauge transformations
(1.12). What happens is that $\varepsilon$ is expanded in
harmonics on the two-sphere according to the relation
\begin{equation}
\varepsilon(t,r,\theta,\varphi)=\sum_{l=0}^{\infty}
\sum_{m=-l}^{l} \varepsilon_{l}(r) Y_{lm}(\theta,\varphi)
e^{i \omega t} .
\end{equation}
After a gauge transformation, one finds from (1.12)
\begin{equation}
[{ }^{\varepsilon}A_{t}]_{\partial M}-[A_{t}]_{\partial M}
=[\nabla_{t}\varepsilon]_{\partial M} ,
\end{equation}
\begin{equation}
[{ }^{\varepsilon}A_{\theta}]_{\partial M}
-[A_{\theta}]_{\partial M}
=[\nabla_{\theta}\varepsilon]_{\partial M},
\end{equation}
\begin{equation}
[{ }^{\varepsilon}A_{\varphi}]_{\partial M}
-[A_{\varphi}]_{\partial M}
=[\nabla_{\varphi}\varepsilon]_{\partial M},
\end{equation}
and by virtue of (1.17) one has
\begin{equation}
[\nabla_{t}\varepsilon]_{\partial M}
=i \omega [\varepsilon]_{\partial M},
\end{equation}
\begin{equation}
[\nabla_{\theta}\varepsilon]_{\partial M}
=\left[\sum_{l=0}^{\infty} \sum_{m=-l}^{l} \varepsilon_{l}(r)
Y_{lm,\theta}(\theta,\varphi)e^{i\omega t}\right]_{\partial M},
\end{equation}
\begin{equation}
[\nabla_{\varphi}\varepsilon]_{\partial M}
=\left[\sum_{l=0}^{\infty} \sum_{m=-l}^{l} \varepsilon_{l}(r)
Y_{lm,\varphi}(\theta,\varphi)e^{i\omega t}\right]_{\partial M}.
\end{equation}
Thus, if $\varepsilon_{l}(r)$ is set to zero at the boundary 
$\forall l$, the right-hand sides of (1.21)--(1.23) vanish at
$\partial M$. But
$$
[\varepsilon_{l}(r)]_{\partial M}=0 \; \forall l
$$
is precisely the condition which ensures the vanishing of
$\varepsilon$ at $\partial M$, and the proof of our statement
is completed.

At this stage, the only boundary condition on
$A_{r}$ whose preservation under the transformation (1.12)
is again guaranteed by Eq. (1.16) is the vanishing of the
gauge-averaging functional, say $\Phi(A)$, at the 
boundary:$^7$
\begin{equation}
[\Phi(A)]_{\partial M}=0 .
\end{equation}
On choosing the Lorenz term $\Phi_{L}(A) \equiv
\nabla^{b}A_{b}$, Eq. (1.24) leads to 
\begin{equation}
{\left[{\partial \over \partial r}(r^{2}A_{r})
\right]}_{\partial M}=0 .
\end{equation}
It is only upon considering the joint effect of Eqs. 
(1.13)--(1.16), (1.24) and (1.25)  
that the whole set of boundary conditions
becomes gauge-invariant. This scheme is also BRST-invariant
(cf. Ref. 10). More precisely, Eq. (1.16) should be
replaced, in the quantum theory, by
a boundary condition on both the real and the
imaginary part of a complex-valued ghost field, say$^7$
\begin{equation}
[\omega]_{\partial M}=[\psi]_{\partial M}=0 .
\end{equation}
Even this terminology is a bit loose, since ghost fields are
really independent of each other, as stressed by DeWitt.$^{11}$
In the calculations of the following sections, it will be
enough to consider a {\it real-valued gauge function} obeying Eq.
(1.16) at the boundary, and then multiply the resulting
contribution to the zero-point energy by $-2$, bearing in
mind Eq. (1.26) and the fermionic nature of ghost fields.
A similar analysis of boundary conditions for the Casimir
effect can be found in Refs. 12 and 13.
Note also that our boundary conditions on the potential lead to
the fulfillment of all ``perfect conductor" boundary conditions
studied in Ref. 2, which involve tangential components
of the electric field and the normal component of the
magnetic field.

The plan of our paper is as follows. Section 2 derives the form
of basis functions for the electromagnetic potential.
Section 3 presents the eigenvalue conditions for the perturbation
modes occurring in the Casimir calculation in the presence
of spherical symmetry, and describes how
the Casimir energy of a conducting spherical shell can be
obtained from such eigenvalue conditions. 
The eigenvalue equations in the axial gauge are studied in
detail in Sec. 4.
Results and open problems are described in Sec. 5, relevant
details are given in the Appendices.
\section*{2. Basis functions for the electromagnetic potential}
\setcounter{section}{2}
\setcounter{equation}{0}
We know from Sec. 1 that,
in our problem, the electromagnetic potential has a temporal
component $A_{t}$, a normal component $A_{r}$, and tangential
components, say $A_{k}$ (hereafter, 
$k$ refers to $\theta$ and $\varphi$). They can all
be expanded in harmonics on the two-sphere, according to the
standard formulas
\begin{equation}
A_{t}(t,r,\theta,\varphi)=\sum_{l=0}^{\infty}\sum_{m=-l}^{l}
a_{l}(r)Y_{lm}(\theta,\varphi)e^{i\omega t},
\end{equation}
\begin{equation}
A_{r}(t,r,\theta,\varphi)=\sum_{l=0}^{\infty}\sum_{m=-l}^{l}
b_{l}(r)Y_{lm}(\theta,\varphi)e^{i\omega t},
\end{equation}
\begin{equation}
A_{k}(t,r,\theta,\varphi)=\sum_{l=1}^{\infty}\sum_{m=-l}^{l}
\Bigr[c_{l}(r)\partial_{k}Y_{lm}(\theta,\varphi)
+d_{l}(r)\varepsilon_{kp}\partial^{p}Y_{lm}(\theta,\varphi)
\Bigr]e^{i\omega t}.
\end{equation}
This means that we are performing a Fourier analysis 
of the components of the electromagnetic potential.
The two terms in square brackets of Eq. (2.3) refer to
longitudinal and transverse modes, respectively.
On setting $\alpha=1$ in Eq. (1.11), the evaluation of
basis functions for electromagnetic perturbations can be
performed after studying the action of the operator
$-g_{\mu\nu}\Box$ on the components (2.1)--(2.3).
For this purpose, we perform the analytic continuation
$\omega \rightarrow iM$, which makes it possible to express
the basis functions in terms of modified Bessel functions
(of course, one could work equally well with $\omega$,
which leads instead to ordinary Bessel functions).
This can be shown by using the 
formulas (A2)--(A4) of Appendix A when $d=2$. For the 
temporal component, one deals with 
the Laplacian acting on a scalar field on the 
three-dimensional disk:
\begin{equation}
({ }^{(3)}\bigtriangleup A)_{t}={\partial^{2}A_{t} \over
\partial r^{2}}+{2\over r}{\partial A_{t}\over \partial r}
+{1\over r^{2}} ({ }^{(2)}\bigtriangleup A)_{t} ,
\end{equation}
which leads to the eigenvalue equation 
\begin{equation}
\left[{d^{2}\over dr^{2}}+{2\over r}{d\over dr}
-{l(l+1)\over r^{2}}\right]a_{l}=M^{2}a_{l} .
\end{equation}
Hereafter, we omit for simplicity any subscript for $M^{2}$,
and the notation for the modes will make it sufficiently
clear which spectrum is studied.
The solution of Eq. (2.5) which is regular at $r=0$
is thus found to be
\begin{equation}
a_{l}(r)={1\over \sqrt{r}}I_{l+1/2}(Mr),
\end{equation}
up to an unessential multiplicative constant.

The action of ${ }^{(3)} \bigtriangleup$ on 
the component $A_{r}$ normal to
the two-sphere is (from Eq. (A3))
\begin{equation}
({ }^{(3)}\bigtriangleup A)_{r}=
{\partial^{2}A_{r}\over \partial r^{2}}+{2\over r}
{\partial A_{r}\over \partial r}
+{1\over r^{2}}({ }^{(2)}\bigtriangleup A)_{r}
-{2\over r^{2}}A_{r}
-{2\over r^{3}}A_{p}^{\; \; \mid p} ,
\end{equation}
where the stroke $\mid$ denotes two-dimensional covariant
differentiation on a two-sphere of unit radius. Last, the
Laplacian on tangential components takes the 
form (cf. Eq. (A4))
\begin{equation}
({ }^{(3)}\bigtriangleup A)_{k}=
{\partial^{2}A_{k}\over \partial r^{2}}
+{1\over r^{2}}({ }^{(2)}\bigtriangleup A)_{k}
-{1\over r^{2}}A_{k}+{2\over r}\partial_{k}A_{r} .
\end{equation}
By virtue of the expansions (2.2) and (2.3), jointly with
the properties (A5) and (A6) of spherical harmonics, Eqs.
(2.7) and (2.8) lead to the eigenvalue equation
\begin{equation}
\left[{d^{2}\over dr^{2}}-{l(l+1)\over r^{2}}
\right]d_{l}=M^{2}d_{l}
\end{equation}
for transverse modes, jointly with entangled eigenvalue
equations for normal and longitudinal modes
(here $l \geq 1$):
\begin{equation}
\left[{d^{2}\over dr^{2}}+{2\over r}{d\over dr}
-{(l(l+1)+2)\over r^{2}}\right]b_{l}
+{2l(l+1)\over r^{3}}c_{l}=M^{2} b_{l},
\end{equation}
\begin{equation}
\left[{d^{2}\over dr^{2}}-{l(l+1)\over r^{2}}
\right]c_{l}+{2\over r}b_{l}=M^{2} c_{l}.
\end{equation}
The mode $b_{0}(r)$ is instead decoupled, and is proportional
to ${I_{3/2}(Mr)\over \sqrt{r}}$ in the interior problem.
It is indeed well known that gauge modes of the Maxwell
field obey a coupled set of eigenvalue equations. In
arbitrary gauges, one cannot decouple these modes.$^{14}$ This
can be proved by trying to put in diagonal form the
$2 \times 2$ operator matrix acting on the modes $b_{l}$
and $c_{l}$ (cf. Refs. 7 and 15). 
In our problem, however, 
with our choice of gauge-averaging functional and gauge
parameter, gauge modes can be disentangled, and a simpler
method to achieve this exists.$^{16}$ For this purpose, we point
out that, since the background is flat, if gauge modes can
be decoupled, they can only reduce to linear combinations 
of Bessel functions, i.e.,
\begin{equation}
b_{l}(r)={B_{\nu}(Mr)\over \sqrt{r}},
\end{equation}
and
\begin{equation}
c_{l}(r)=C(\nu)B_{\nu}(Mr)\sqrt{r} .
\end{equation}
With our notation, $C(\nu)$ is some constant depending on
$\nu$, which is obtained in turn from $l$. To find $\nu$ 
and $C(\nu)$, we insert the ansatz (2.12) and (2.13) into the
system of equations (2.10) and (2.11), and we require that 
the resulting equations should be of Bessel type for 
$B_{\nu}(Mr)$, i.e.,
\begin{equation}
B_{\nu}''+{B_{\nu}' \over Mr}
-{\nu^{2}\over M^{2}r^{2}}B_{\nu} 
-M^{2}B_{\nu}=0 .
\end{equation}
This leads to two algebraic equations for $\nu^{2}$:
\begin{equation}
\nu^{2}={9\over 4}+l(l+1)-2l(l+1)C ,
\end{equation}
and
\begin{equation}
\nu^{2}={1\over 4}+l(l+1)-{2\over C}.
\end{equation}
By comparison, one thus finds a second-order algebraic
equation for C:
\begin{equation}
l(l+1)C^{2}-C-1=0,
\end{equation}
whose roots are
\begin{equation}
C_{+}={1\over l},
\end{equation}
and
\begin{equation}
C_{-}=-{1\over (l+1)}.
\end{equation}
The corresponding values of $\nu$ are
\begin{equation}
\nu_{+}=l-{1\over 2},
\end{equation}
and
\begin{equation}
\nu_{-}=l+{3\over 2} .
\end{equation}
Hence one finds the basis functions for normal and 
longitudinal perturbations in the interior problem in the form
\begin{equation}
b_{l}(r)=\alpha_{1,l}{I_{l+3/2}(Mr)\over \sqrt{r}}
+\alpha_{2,l}{I_{l-1/2}(Mr)\over \sqrt{r}},
\end{equation}
\begin{equation}
c_{l}(r)=-{\alpha_{1,l}\over (l+1)}
I_{l+3/2}(Mr)\sqrt{r}+{\alpha_{2,l}\over l}
I_{l-1/2}(Mr)\sqrt{r},
\end{equation}
whereas, from Eq. (2.9), transverse modes read
\begin{equation}
d_{l}(r)=I_{l+1/2}(Mr)\sqrt{r}.
\end{equation}
Last, but not least, ghost modes obey an eigenvalue equation 
analogous to (2.5) (see comments after (1.26)), i.e.
\begin{equation}
\left[{d^{2}\over dr^{2}}+{2\over r}{d\over dr}
-{l(l+1)\over r^{2}}\right]\varepsilon_{l}=M^{2}\varepsilon_{l},
\end{equation}
and hence they read
\begin{equation}
\varepsilon_{l}(r)={1\over \sqrt{r}}I_{l+1/2}(Mr).
\end{equation}

In the exterior problem, i.e., for $r$ greater than the two-sphere
radius $R$, one has simply to replace the modified Bessel functions
of first kind in Eqs. (2.6), (2.22)--(2.24) 
and (2.26) by modified Bessel
functions of second kind, to ensure regularity at infinity.
\section*{3. Eigenvalue conditions and Casimir energy}
\setcounter{section}{3}
\setcounter{equation}{0}
In our problem, ghost modes are of course decoupled from the 
modes for the electromagnetic potential which occur in the 
expansions (2.1)--(2.3). Nevertheless, this does not mean that
they do not play a role in the Casimir energy calculation. By
contrast, we already know from Sec. 1 (see comments after (1.16))
that boundary conditions on the ghost are strictly necessary to
ensure gauge invariance of the boundary conditions on the potential. 
It is then clear that such boundary conditions, combined with the
differential equation (2.25), lead to a ghost spectrum whose
contribution to the Casimir energy can only be obtained after a
detailed calculation (e.g. Green-function approach, or
$\zeta$-function regularization). As is pointed out in Ref. 12,
it is not surprising that ghost terms are important, since one
has already included effects of other degrees of freedom which
should be compensated for. This issue is further clarified
by the analysis of the original Casimir problem: two perfectly
conducting parallel plates. In a covariant formalism, one has to
consider the energy-momentum tensor for ghosts, which is found to
give a non-vanishing contribution to the average energy density
in vacuum (see Sec. 4.5 of Ref. 7). After taking into account 
boundary conditions for the potential entirely analogous to our
Eqs. (1.13)--(1.15) and (1.24), one then finds a renormalized value
of the zero-point energy in complete agreement with the result first
found by Casimir.$^1$

By virtue of Eq. (1.13), the modes $a_{l}(r)$ obey homogeneous
Dirichlet conditions:
\begin{equation}
[a_{l}(r)]_{\partial M}=0, \forall l \geq 0.
\end{equation}
Moreover, Eq. (1.25) implies that the modes $b_{l}$ obey the
boundary conditions
\begin{equation}
\left[{\partial \over \partial r} r^{2}b_{l}(r)
\right]_{\partial M}=0, \forall l \geq 0.
\end{equation}
Last, the modes $c_{l}$ and $d_{l}$, being the tangential
modes, obey Dirichlet boundary conditions (cf. Eqs. (1.14)
and (1.15))
\begin{equation}
[c_{l}(r)]_{\partial M}=[d_{l}(r)]_{\partial M}=0,
\forall l \geq 1.
\end{equation}
On taking into account how the modes are expressed in
terms of Bessel functions (see Eqs. (2.6), (2.22)--(2.24)),
one thus finds five sets of eigenvalue conditions
for the interior and exterior problems, respectively:
\vskip 0.3cm
\noindent
(i) Temporal modes: 
\begin{mathletters}
\begin{equation}
I_{l+1/2}(MR)=0, \forall l \geq 0,
\end{equation}
\begin{equation}
K_{l+1/2}(MR)=0, \forall l \geq 0.
\end{equation}
\end{mathletters}
\vskip 0.3cm
\noindent
(ii) Decoupled normal mode:
\begin{mathletters}
\begin{equation}
\left[{\partial \over \partial r}r^{3/2}I_{3/2}(Mr)
\right]_{r=R}=0,
\end{equation}
\begin{equation}
\left[{\partial \over \partial r}r^{3/2}K_{3/2}(Mr)
\right]_{r=R}=0.
\end{equation}
\end{mathletters}
\vskip 0.3cm
\noindent
(iii) Coupled longitudinal and normal modes (here
$\nu \equiv l + 3/2$):
\begin{mathletters}
\begin{eqnarray}
\; & \; &(\nu-1/2)I_{\nu}'(MR)I_{\nu -2}(MR)
+3(\nu-1){I_{\nu}(MR)\over MR}I_{\nu-2}(MR) \nonumber \\
&+& (\nu-3/2)I_{\nu-2}'(MR)I_{\nu}(MR)=0,
\end{eqnarray}
\begin{eqnarray}
\; & \; & (\nu-1/2)K_{\nu}'(MR)K_{\nu -2}(MR)
+3(\nu-1){K_{\nu}(MR)\over MR}K_{\nu-2}(MR) \nonumber \\
&+& (\nu-3/2)K_{\nu-2}'(MR)K_{\nu}(MR)=0.
\end{eqnarray}
\end{mathletters}
\vskip 0.3cm
\noindent
(iv) Transverse modes:
\begin{mathletters}
\begin{equation}
I_{l+1/2}(MR)=0, \forall l \geq 1,
\end{equation}
\begin{equation}
K_{l+1/2}(MR)=0, \forall l \geq 1.
\end{equation}
\end{mathletters}
Note that, the boundary being a two-sphere, our ``transverse
photons" are two-dimensional, unlike other physical problems
where transverse photons are three-dimensional since the
boundary is a three-sphere (or another three-surface).
\vskip 0.3cm
\noindent
(v) Ghost modes (multiplying their $\zeta$-function by -2):
\begin{mathletters}
\begin{equation}
I_{l+1/2}(MR)=0, \forall l \geq 0,
\end{equation}
\begin{equation}
K_{l+1/2}(MR)=0, \forall l \geq 0.
\end{equation}
\end{mathletters}

The eigenvalue conditions (3.5a) and (3.5b) can be re-expressed
in the form
\begin{equation}
I_{1/2}(MR)=0,
\end{equation}
\begin{equation}
K_{1/2}(MR)=0.
\end{equation}
It is thus clear that, by construction, the contribution of Eqs.
(3.9) and (3.10) to the Casimir energy of a conducting spherical
shell cancels exactly the joint effect of Eqs. (3.4a), (3.4b),
(3.7a), (3.7b), (3.8a) and (3.8b), bearing in mind the fermionic
nature of ghost fields. In general, each set of boundary conditions
involving a set of positive eigenvalues, say $\left \{ \lambda_{k}
\right \}$, contributes to the Casimir energy in a way which is
clarified by the $\zeta$-function method, because the regularized
ground-state energy is defined by the equation (for ${\text {Re}}(s)
> s_{0}=2$)
\begin{equation}
E_{0}(s) \equiv -{1\over 2} \sum_{k} (\lambda_{k})^{{1\over 2}-s}
\; \mu^{2s}=-{1\over 2} \zeta \left(s-{1\over 2} \right)\mu^{2s},
\end{equation}
which is later analytically continued to the value $s=0$ in the
complex-$s$ plane. Here $\mu$ is the usual mass parameter and $\zeta$
is the $\zeta$-function of the positive-definite elliptic operator 
$\cal B$ with discrete spectrum $\left \{ \lambda_{k} \right \}$:
\begin{equation}
\zeta_{{\cal B}}(s) \equiv {\text {Tr}}_{L^{2}}({\cal B}^{-s})
=\sum_{k} (\lambda_{k})^{-s}.
\end{equation}
In other words, the regularized ground-state energy is equal to
$-{1\over 2}\zeta_{{\cal B}}(-1/2)$. Although the eigenvalues are
known only implicitly, the form of the function occurring in the
mode-by-mode expression of the boundary conditions leads eventually
to $E_{0}(0)$.$^{6,8}$

The non-trivial part of the analysis is represented by Eqs. (3.6a)
and (3.6b). These are obtained by imposing the Robin boundary
conditions for normal modes, and Dirichlet conditions for 
longitudinal modes. To find nontrivial solutions of the resulting
linear systems of two equations in the unknowns $\alpha_{1,l}$
and $\alpha_{2,l}$, the determinants of the matrices of coefficients
should vanish. This leads to Eqs. (3.6a) and (3.6b). At this stage,
it is more convenient to re-express such equations in terms of 
Bessel functions of order $l+1/2$. On using the standard recurrence
relations among Bessel functions and their first derivatives, one
thus finds the following equivalent forms of eigenvalue conditions:
\begin{equation}
I_{l+1/2}(MR)\left[I_{l+1/2}'(MR)+{1\over 2MR}
I_{l+1/2}(MR)\right]=0, \forall l \geq 1,
\end{equation}
\begin{equation}
K_{l+1/2}(MR)\left[K_{l+1/2}'(MR)+{1\over 2MR}
K_{l+1/2}(MR)\right]=0, \forall l \geq 1.
\end{equation}
Thus, the contribution of the coupled normal and longitudinal
modes splits into the sum of contributions of two scalar fields
obeying Dirichlet and Robin boundary conditions, respectively,
with the $l=0$ mode omitted. This corresponds exactly to the
contributions of TE and TM modes (Eqs. (1.9) and (1.10)), and
gives the same contribution as the one found by Boyer.$^2$ 
\section*{4. Axial gauge}
\setcounter{section}{4}
\setcounter{equation}{0}
As is well known, on using the Faddeev-Popov
formalism, one performs Gaussian averages over gauge
functionals, say $\chi^{\mu}$, by adding to the original 
Lagrangian a gauge-averaging term 
$\chi^{\mu}\beta_{\mu \nu}\chi^{\nu}$, where $\beta_{\mu \nu}$
is any constant invertible matrix.$^{11}$
This has the effect of turning 
the original operator on field perturbations into a new operator
which has the advantage of being non-degenerate. It can also
become of Laplace type for suitable choices of some gauge
parameters. Moreover, the result of a one-loop calculation is
expected to be $\chi$- {\it and} $\beta$-independent, although
no rigorous proof exists on manifolds with boundary.$^7$ In
particular, for Maxwell theory, $\beta_{\mu \nu}$
reduces to a $1 \times 1$ matrix, i.e. a real-valued parameter,
and $\chi^{\mu}$ reduces to the familiar covariant or 
noncovariant gauges for Maxwell theory, e.g. Lorenz, Coulomb,
axial, ... . It is therefore quite important to understand the
key features of the Casimir-energy calculations in all such gauges.
For this purpose, we consider a gauge-averaging functional of the
axial type, i.e.
\begin{equation}
\Phi(A) \equiv n^{\mu}A_{\mu},
\end{equation}
where $n^{\mu}$ is the unit normal vector field  
$n^{\mu}=(0,1,0,0)$. The resulting gauge-field operator is
found to be, in our flat background,
\begin{equation}
P^{\mu \nu}=-g^{\mu \nu}\Box + \nabla^{\mu} \nabla^{\nu}
+{1\over \alpha}n^{\mu}n^{\nu}.
\end{equation}
Note that, unlike the case of Lorenz gauge, the $\alpha$
parameter is dimensionful and has dimension $[\text{length}]^{2}$
(see Ref. 17).

Now we impose again the boundary condition according to which the
gauge-averaging functional should vanish at $\partial M$:
\begin{equation}
[\Phi(A)]_{\partial M}=[n^{\mu}A_{\mu}]_{\partial M}
=[A_{r}]_{\partial M}=0,
\end{equation}
which implies that all $b_{l}$ modes vanish at the boundary
(cf. Eq. (3.2)). 

A further consequence of the axial gauge may be derived by acting
on the field equations 
\begin{equation}
P^{\mu \nu}A_{\nu}=0
\end{equation}
with the operations of covariant differentiation and contraction
with the unit normal, i.e. 
\begin{equation}
\nabla_{\mu}P^{\mu \nu}A_{\nu}=0,
\end{equation}
\begin{equation}
n_{\mu}P^{\mu \nu}A_{\nu}=0.
\end{equation}
Eq. (4.5) leads, in flat space, to the differential equation
\begin{equation}
{\partial A_{r}\over \partial r}+{2\over r}A_{r}=0.
\end{equation}
This first-order equation leads to $b_{l}$ modes having the form
\begin{equation}
b_{l}={b_{0,l}\over r^{2}}.
\end{equation}
Thus, by virtue of the boundary condition (4.3), the modes
$b_{l}$ vanish everywhere, and hence $A_{r}$ vanishes
identically if the axial gauge-averaging functional is chosen with
such boundary conditions.

Moreover, Eq. (4.6) leads to the equation
\begin{equation}
-i\omega {\partial A_{t}\over \partial r}
+{1\over r^{2}}{\partial \over \partial r}
A_{k}^{\; \mid k}=0,
\end{equation}
which implies, upon making the analytic continuation
$\omega \rightarrow iM$, 
\begin{equation}
M{da_{l}\over dr}-{l(l+1)\over r^{2}}{dc_{l}\over dr}=0.
\end{equation}
At this stage, the transverse modes $d_{l}$ obey again the
eigenvalue equation (2.9), whereas the remaining set of modes
obey differential equations which, from Eq. (4.4), are found
to be
\begin{equation}
\left[{d^{2}\over dr^{2}}+{2\over r}{d\over dr}
-{l(l+1)\over r^{2}}\right]a_{l}-M{l(l+1)\over r^{2}}c_{l}=0,
\end{equation}
\begin{equation}
{d^{2}c_{l}\over dr^{2}}-Ma_{l}-M^{2}c_{l}=0,
\end{equation}
\begin{equation}
a_{0}=0.
\end{equation}
In particular, Eq. (4.13) is obtained from Eq. (4.12)
(when $l=0$), which is
a reduced form of the Eq. $P_{k\nu}A^{\nu}=0$ upon bearing in
mind that $b_{l}$ modes vanish everywhere. 

Last, but not least, one should consider the ghost operator, which,
in the axial gauge, is found to be 
\begin{equation}
Q=-{\partial \over \partial r}.
\end{equation}
This leads to ghost modes having the form
\begin{equation}
\varepsilon_{l}=\varepsilon_{0,l} \; e^{-Mr}.
\end{equation}
On the other hand, following the arguments developed in Sec. 1,
one can prove that, also in the axial gauge, the ghost field
should vanish at the boundary, to ensure gauge invariance of the
whole set of boundary conditions on the potential. It is then
clear, from Eq. (4.15), that ghost modes vanish everywhere in
the axial gauge.

On studying the system (4.10)--(4.12) one has first to prove 
that these three equations are compatible. This is indeed the
case, because differentiation with respect to $r$ of Eq. (4.10)
leads to a second-order equation which, upon expressing
${dc_{l}\over dr}$ from Eq. (4.10) and ${d^{2}c_{l}\over dr^{2}}$
from Eq. (4.12), is found to coincide with Eq. (4.11). 
Thus, we have a system of two second-order differential
equations for two functions $a_{l}$ and $c_{l}$. However, these
functions are not independent, in that they are connected by
Eq. (4.10). Hence for every value of $l$ one has one degree of
freedom instead of two. Finally, we have for every $l$ two degrees
of freedom, one resulting from Eqs. (4.10)--(4.12), and another, i.e.
the transverse mode $d_{l}$. 
Thus, an estimate of the number of degrees of freedom giving
contribution to Casimir energy coincides with that in other gauges.
Moreover, the parameter $\alpha$ does not affect the Casimir 
energy, since $\alpha$ does not occur in any of the eigenvalue
equations.
 
Unfortunately, we cannot obtain the exact form of the solutions
of Eqs. (4.11) and (4.12) in terms of special functions
(e.g. Bessel or hypergeometric). This crucial point can be made
precise by remarking that, if it were possible to disentangle
the system (4.11) and (4.12), one could find some functions 
$\alpha_{l}, \beta_{l}, V_{l}, W_{l}$, say, such that the
$2 \times 2$ matrix 
$$
\pmatrix{1 & V_{l} \cr W_{l} & 1 \cr}
\pmatrix{{\hat A}_{l} & {\hat B}_{l} \cr
{\hat C}_{l} & {\hat D}_{l} \cr}
\pmatrix{1 & \alpha_{l} \cr \beta_{l} & 1 \cr}
$$
has no non-vanishing off-diagonal elements, where the operators
${\hat A}_{l}, {\hat B}_{l}, {\hat C}_{l}, {\hat D}_{l}$ are the
ones occurring in Eqs. (4.11) and (4.12) (cf. Refs. 7,14). For
example, the first off-diagonal element of such matrix is the 
operator 
$$
{\hat A}_{l} \alpha_{l}+{\hat B}_{l}+V_{l}({\hat C}_{l}\alpha_{l}
+{\hat D}_{l}).
$$
On setting to zero the coefficients of ${d^{2}\over dr^{2}}$ and
${d\over dr}$ one finds that $\alpha_{l}=-V_{l}
={\alpha_{0,l}\over r}$, where $\alpha_{0,l}$ is a constant. But
it is then impossible to set to zero the ``potential" term of this
operator, i.e. its purely multiplicative part.

Nevertheless, there is some evidence that the axial gauge may be
consistently used to evaluate the Casimir energy. For this
purpose we find it helpful to consider a simpler problem, i.e.
the Casimir energy in the axial gauge for the
case of flat boundary. In this case the basis functions are plane
waves
\begin{equation}
A_{\mu}=A_{0,\mu}e^{i(k_{x}x+k_{y}y+k_{z}z-\omega t)},
\end{equation}  
where the admissible values of $k_{x}, k_{y}, k_{z}$ are
determined by the boundary conditions, which are here taken
to be analogous to the case of a curved boundary. Let us choose
the conducting boundaries parallel to the $x$- and $y$-axes,
while the vector $n^{\mu}$ is directed along the $z$-axis.
Then the condition of compatibility of field equations in the
axial gauge is reduced to 
\begin{equation}
D \equiv {\text {det}} 
\pmatrix{-k^{2} & \omega k_{x} & \omega k_{y} & \omega k_{z} \cr
\omega k_{x} & k_{y}^{2}+k_{z}^{2}-\omega^{2} &
-k_{x} k_{y} & -k_{x} k_{z} \cr
\omega k_{y} & -k_{x} k_{y} & k_{x}^{2}+k_{z}^{2}-\omega^{2}
& -k_{y} k_{z} \cr
\omega k_{z} & -k_{x}k_{z} & -k_{y} k_{z} 
& k_{x}^{2}+k_{y}^{2}-\omega^{2}+{1\over \alpha} \cr}=0.
\end{equation}
Direct calculation shows that this determinant is equal to
\begin{equation}
D=-{1\over \alpha}k_{z}^{2}\left(\omega^{2}-k^{2}\right)^{2}.
\end{equation}
Hence we have reproduced the correct dispersion relation
between energy $\omega$ and wave number $k$ (to every admissible
value of $k$ there correspond two contributions to the Casimir
energy of the form $\omega = \mid {\vec k} \mid$). Of course,
no non-vanishing ghost modes exist, once that the axial type
gauge-averaging functional is set to zero at the boundary
(cf. (4.3), (4.14) and (4.15)). Interestingly, on imposing
the boundary conditions, one can set to zero the $A_{z}$
component of the electromagnetic potential as was done with
$A_{r}$ in the spherical case. In this case the compatibility
condition of field equations is reduced to the vanishing of the
determinant of a $3 \times 3$ matrix, obtained by omitting the
fourth row and the fourth column in Eq. (4.17). One can easily
see that the determinant of this $3 \times 3$ matrix coincides
with that of the $4 \times 4$ matrix up to a multiplicative
factor ${1\over \alpha}$, which does not affect the dispersion
relation. 

To sum up, a complete correspondence can be established between
the key features of the axial gauge in the cases of flat and
curved boundary: the $\alpha$ parameter does not affect the Casimir
energy, the component of the potential orthogonal to the boundary
vanishes, ghost modes vanish, and only two independent degrees
of freedom contribute. On going from the flat to the curved case,
however, the analysis of the dispersion relation is replaced by
the problem of finding explicit solutions of Eqs. (4.10)--(4.12),
with the corresponding eigenvalue conditions. This last technical
problem goes beyond the present capabilities of the authors, but
the exact results and complete correspondences established so far
seem to add evidence in favour of a complete solution being in
sight in the near future.

\section*{5. Concluding remarks}
\setcounter{section}{5}
\setcounter{equation}{0}
Scientific knowledge makes progress not only when an entirely
new result is first obtained, but also when alternative derivations
of such properties become available, possibly showing that further
important problems are in sight. The contributions of our paper
fall within the latter framework. For this purpose, we have
studied an approach to the evaluation of the
zero-point energy of a conducting spherical shell which relies
on a careful investigation of the potential and of ghost fields,
with the corresponding set of boundary conditions on $A_{\mu}$
perturbations and ghost modes. When Boyer first developed his
calculation,$^2$ the formalism of ghost fields for the quantization
of gauge fields and gravitation had just 
been developed,$^{18-20}$ and
hence it is quite natural that, in the first series of papers on
Casimir energies, the ghost contribution was not considered, 
since the emphasis was always put on TE and TM modes for the
electromagnetic field. On the other hand, the Casimir energy is a
peculiar property of the quantum theory, and an approach via
path-integral quantization regards the potential and the ghost as
more fundamental.$^{7,11}$ This is indeed necessary to appreciate
that Maxwell theory is a gauge theory. 
Some of these issues were studied in 
Refs. 7,12,13,21,22, including calculations with ghosts 
in covariant gauges, 
but, to our knowledge, an explicit mode-by-mode analysis of the
$A_{\mu}$ and ghost contributions in problems with spherical
symmetry was still lacking in the 
literature. The contributions of our investigation are as follows.

First, the basis functions for the electromagnetic potential have
been found explicitly when the Lorenz gauge-averaging functional
is used. The temporal, normal, longitudinal and transverse modes
have been shown to be linear combinations of spherical Bessel
functions. Second, it has been proved that transverse modes of the
potential are, by themselves, unable to
reproduce the correct value for the Casimir energy of a conducting
spherical shell. Third, it is exactly the effect of coupled 
longitudinal and normal modes of $A_{\mu}$ which is responsible for
the value of $\bigtriangleup E$ given in Eq. (1.2), following
Boyer$^2$ (see also Refs. 3,4,6). This adds evidence in favour 
of physical degrees of freedom for gauge theories being a concept
crucially depending on the particular problem under consideration
and on the boundary conditions. Further study is necessary to
understand this property.$^{14,21}$ Fourth, ghost modes play a
nontrivial role as well, in that they cancel the contribution 
resulting from transverse, decoupled and temporal 
modes of the potential. Fifth, the axial gauge-averaging functional
has been used to study the Casimir energy for Boyer's problem.
Unlike the case of the Lorenz gauge, ghost modes and normal
modes are found to vanish, and one easily proves that the result
is independent of the $\alpha$ parameter. A complete comparison
with the case of flat boundary has also been performed, getting 
insight into the problems of independent degrees of freedom
and non-vanishing contributions to the Casimir energy. 

Indeed, recent investigations of Euclidean Maxwell theory in 
quantum cosmological backgrounds had already shown that
longitudinal, normal and ghost modes are all essential to
obtain the value of the conformal anomaly and of the
one-loop effective action.$^{7,14,15,23}$ In other words, the
mode-by-mode evaluation of the Faddeev--Popov path integral
in the one-loop approximation agrees with the result 
obtained, for the same path integral, by using the method
of conformal variations with the corresponding Schwinger--DeWitt
asymptotic expansion.$^{7,11}$ For example, in a
portion of flat Euclidean four-space bounded by two concentric
three-spheres, the contribution to $\zeta(0)$ of the decoupled
normal mode cancels exactly the one resulting from transverse
modes. Moreover, on the four-dimensional disk, transverse modes
give contributions of opposite signs to $\zeta(0)$ with 
magnetic or electric boundary conditions,$^7$ whereas the 
$\zeta(0)$ value is the same for both boundary conditions and
equal to $-31/90$.$^{14,15}$ Thus, in problems where the background
is four-dimensional and the boundary is three-dimensional, 
the covariant $\zeta$-function calculations,
including ghost modes, give results in
disagreement with a calculation relying on boundary conditions
for the fields and reduction to physical degrees of
freedom.$^{7,14,15,24}$ Further evidence of the
nontrivial role played by ghost modes in curved backgrounds
had been obtained, much earlier, in Ref. 25. Hence, we find it
nontrivial that the ghost formalism 
gives results in complete agreement with Boyer's investigation.$^2$
Note also that, in the case of
perfectly conducting parallel plates, the ghost contribution
cancels the one due to tangential components of the potential
(see Sec. 4.5 of Ref. 7). This differs from the cancellations
found in our paper in the presence of spherical symmetry.

The main open problem is now the explicit proof that the Casimir
energy is independent of the choice of
gauge-averaging functional. This task is as difficult as
crucial to obtain a thorough understanding of the quantized
electromagnetic field in the presence of bounding surfaces.
In general, as shown in Sec. 4, 
one has then to study entangled eigenvalue
equations for temporal, normal and longitudinal modes.
The general solution is not (obviously)
expressed in terms of well known special functions (cf., for
example, Ref. 14). A satisfactory understanding of this 
problem is still lacking, while its solution would be of great
relevance both for the foundations and for the applications of
quantum field theory. In covariant gauges, coupled eigenvalue
equations with arbitrary gauge parameter also lead to severe
technical problems, which are described in Appendix B.
A further set of nontrivial applications 
lies in the analysis of more involved geometries, where spherical
symmetry no longer holds, or in the investigation of media which
are not perfect conductors.$^{26-28}$ Thus, there exists increasing
evidence that the study of Casimir energies will continue to play
an important role in quantum field theory in the years to come,
and that a formalism relying on potentials and ghost fields is,
indeed, crucial on the way
towards a better understanding of quantized fields.

\acknowledgements
A.K. was partially supported by RFBR via Grant No. 
96-02-16220, and is grateful to CARIPLO Science Foundation
for financial support.
K.K. is indebted to the Istituto Nazionale di Fisica Nucleare
for financial support and to the Dipartimento di Scienze Fisiche
of the University of Naples for hospitality. This investigation
has been partly supported by the DFG under contract 
number BO1112/4-2. The authors are grateful to Michael Bordag
and Dmitri Vassilevich for enlightening conversations.

\appendix
\section{}
One can consider, in general, a $(d+1)$-dimensional disk
with metric
\begin{equation}
g=dr \otimes dr + r^{2} \Omega_{d} , 
\; r \in [0,R],
\end{equation}
where $\Omega_{d}$ is the metric on the $d$-dimensional sphere
of unit radius. The action of the Laplace operator on a scalar
field, say $\phi$, is found to be
\begin{equation}
\bigtriangleup \phi=\left({\partial^{2}\over \partial r^{2}}
+{d\over r}{\partial \over \partial r}
+{1\over r^{2}} { }^{(d)} \bigtriangleup \right) \phi ,
\end{equation}
whereas the action of the same operator on normal and tangential
components of a one-form is 
\begin{equation}
(\bigtriangleup A)_{r}=\left({\partial^{2}\over \partial r^{2}}
+{d\over r}{\partial \over \partial r}
+{1\over r^{2}} { }^{(d)}\bigtriangleup 
-{d\over r^{2}} \right)A_{r}
-{2\over r^{3}}A_{p}^{\; \; \mid p} ,
\end{equation}
\begin{equation}
(\bigtriangleup A)_{k}=\left({\partial^{2}\over \partial r^{2}}
+{(d-2)\over r}{\partial \over \partial r}
+{1\over r^{2}} { }^{(d)}\bigtriangleup 
-{(d-1)\over r^{2}}\right)A_{k}
+{2\over r}\partial_{k}A_{r}.
\end{equation}
With our notation, ${ }^{(d)}\bigtriangleup \equiv
_{\mid p}^{\; \; \; \mid p}$ is the Laplacian on a $d$-sphere
of unit radius. When $d=2$, its action on spherical harmonics
is such that
\begin{equation}
{ }^{(2)}\bigtriangleup Y_{lm}(\theta, \varphi)
=-l(l+1) Y_{lm}(\theta, \varphi) .
\end{equation}
Another useful formula in the course of deriving Eq. (2.11) 
from Eq. (2.8) is the one according to which$^{29}$
\begin{equation}
{ }^{(2)}\bigtriangleup \partial_{k} Y_{lm}(\theta, \varphi)
=[-l(l+1)+1] \partial_{k} Y_{lm}(\theta, \varphi) .
\end{equation}

\section{}
A naturally occurring question is whether one
can prove explicitly that the Casimir energy is independent
of the value of the $\alpha$ parameter in the operator (1.11).
Indeed, from Eqs. (2.4), (2.7), (2.8), jointly with the form
of the Lorenz gauge-averaging functional:
\begin{equation}
\Phi_{L}(A)=-{\partial A_{t}\over \partial t}
+{\partial A_{r}\over \partial r}
+{1\over r^{2}}A_{k}^{\; \mid k}
+{2\over r}A_{r},
\end{equation}
one finds that
\begin{equation}
-P_{t \nu}A^{\nu}=\left({\omega^{2}\over \alpha} 
+\bigtriangleup \right)A_{t}
+i\omega \left({1\over \alpha}-1 \right)
\left({\partial A_{r}\over \partial r}
+{1\over r^{2}}A_{k}^{\; \mid k}
+{2\over r}A_{r}\right),
\end{equation}
\begin{equation}
-P_{r \nu}A^{\nu}=(\omega^{2}+\bigtriangleup)A_{r}
+i \omega \left(1-{1\over \alpha}\right)
{\partial A_{t}\over \partial r}
-\left(1-{1\over \alpha}\right){\partial \over \partial r}
\left({\partial A_{r}\over \partial r}
+{1\over r^{2}}A_{k}^{\; \mid k}
+{2\over r}A_{r}\right),
\end{equation}
\begin{equation}
-P_{k \nu}A^{\nu}=(\omega^{2}+\bigtriangleup)A_{k}
+i \omega \left(1-{1\over \alpha}\right)
{\partial A_{t}\over \partial x^{k}}
-\left(1-{1\over \alpha}\right){\partial \over \partial x^{k}}
\left({\partial A_{r}\over \partial r}
+{1\over r^{2}}A_{p}^{\; \mid p}
+{2\over r}A_{r}\right).
\end{equation}
As in Sec. 2, we now perform the analytic continuation
$\omega \rightarrow iM$. For arbitrary values of $\alpha$, this
remains a well defined procedure, but it no longer leads to a
system admitting Bessel type solutions. By contrast, by virtue of
the equations (2.1)--(2.3) and (B.2)--(B.4), one finds a coupled
system of three ordinary differential equations with variable
coefficients. On defining the column vector
\begin{equation}
u \equiv \pmatrix{a_{l} \cr b_{l} \cr c_{l} \cr},
\end{equation}
jointly with the nine operators
\begin{equation}
A_{11} \equiv {d^{2}\over dr^{2}}+{2\over r}{d\over dr}
-{l(l+1)\over r^{2}}+\left(1-{1\over \alpha} \right)M^{2},
\end{equation}
\begin{equation}
A_{12} \equiv M \left(1-{1\over \alpha}\right)
\left({d\over dr}+{2\over r}\right),
\end{equation}
\begin{equation}
A_{13} \equiv M \left({1\over \alpha}-1 \right)
{l(l+1)\over r^{2}},
\end{equation}
\begin{equation}
A_{21} \equiv M \left({1\over \alpha}-1 \right){d\over dr},
\end{equation}
\begin{equation}
A_{22} \equiv {1\over \alpha}{d^{2}\over dr^{2}}
+{2\over \alpha}{1\over r}{d\over dr}
-\left({2\over \alpha}+l(l+1)\right){1\over r^{2}},
\end{equation}
\begin{equation}
A_{23} \equiv \left(1-{1\over \alpha}\right)
{l(l+1)\over r^{2}}{d\over dr}
+{2\over \alpha}{l(l+1)\over r^{3}},
\end{equation}
\begin{equation}
A_{31} \equiv M \left({1\over \alpha}-1 \right),
\end{equation}
\begin{equation}
A_{32} \equiv \left({1\over \alpha}-1 \right)
{d\over dr}+{2\over \alpha}{1\over r},
\end{equation}
\begin{equation}
A_{33} \equiv {d^{2}\over dr^{2}}
-{l(l+1)\over \alpha r^{2}},
\end{equation}
the resulting system reads (with summation over the repeated
index $p=1,2,3$)
\begin{equation}
A_{jp}u_{p}=M^{2}u_{j},
\end{equation}
for all $j=1,2,3$.
Remarkably, the temporal, normal and longitudinal modes are
now {\it all coupled}, but the system of equations is so
complicated that no explicit solution in terms of well known
special functions can be obtained, as far as the authors 
can see. Nevertheless, if $\alpha \not = 1$, regular 
basis functions still exist, since $r=0$ is a Fuchsian
singularity of the system (B.15). This implies that the
desired solutions can be expressed, in the interior
problem, in the form
$$
r^{\rho}\sum_{k=0}^{\infty}\eta_{l,k}r^{k},
$$
where $\rho$ is a real-valued parameter, and 
$\eta_{l,k}$ are some coefficients.
One should also say that, in the quantum cosmological 
backgrounds studied over the last few years,$^7$ the problem
is less severe, since only normal and longitudinal modes
remain coupled for arbitrary values of $\alpha$ (but in that 
case there is no time variable, the background is 
four-dimensional and the boundary is three-dimensional). Only
the transverse modes are decoupled, and obey the equation (2.9).
Moreover, the modes $a_{0}$ and $b_{0}$ are coupled as well
(unlike the case when $\alpha=1$), and obey the following
system of three eigenvalue equations:
\begin{equation}
\left[{d^{2}\over dr^{2}}+{2\over r}{d\over dr}
+\left(1-{1\over \alpha} \right)M^{2}\right]a_{0}
+M \left(1-{1\over \alpha}\right)
\left({db_{0}\over dr}+{2\over r}b_{0} \right)
=M^{2} a_{0},
\end{equation}
\begin{equation}
\left[{1\over \alpha}{d^{2}\over dr^{2}}
+{2\over \alpha}{1\over r}{d\over dr}
-{2\over \alpha}{1\over r^{2}}\right]b_{0}
+M \left({1\over \alpha}-1 \right){da_{0}\over dr}
=M^{2} b_{0},
\end{equation}
\begin{equation}
M\left({1\over \alpha}-1 \right)a_{0}
+\left({1\over \alpha}-1 \right){db_{0}\over dr}
+{2\over \alpha}{b_{0}\over r}=0.
\end{equation}
Similar technical problems occur on considering other choices
of the gauge-averaging functional, e.g.
$$
\Phi(A) \equiv {\partial A_{r}\over \partial r}
+{1\over r^{2}}A_{k}^{\; \mid k}
+{2\over r}A_{r},
$$
and hence we only write
down the eigenvalue equations (B.15) which are obtained
in the Lorenz case.

Since in Sec. 4 we have found it quite helpful to perform a
comparison with the analysis of a flat boundary, here we
describe the same procedure for the Lorenz gauge. On using
again the plane-wave expansion (4.16) for the electromagnetic
potential, one gets the following compatibility condition
for field equations:
\begin{equation}
F \equiv {\text {det}} \pmatrix{\omega^{2}-k^{2}-\beta \omega^{2}
& \beta \omega k_{x} & \beta \omega k_{y} & \beta \omega k_{z} \cr
\beta \omega k_{x} & k^{2}-\omega^{2}-\beta k_{x}^{2} 
& -\beta k_{x} k_{y} & -\beta k_{x} k_{z} \cr
\beta \omega k_{y} & -\beta k_{x} k_{y} &
k^{2}-\omega^{2}-\beta k_{y}^{2} &
-\beta k_{y} k_{z} \cr
\beta \omega k_{z} & -\beta k_{x} k_{z} & 
-\beta k_{y} k_{z} & k^{2}-\omega^{2}-\beta k_{z}^{2} \cr}=0,
\end{equation}
where $\beta \equiv 1-{1\over \alpha}$. The direct calculation
shows that 
\begin{equation}
F=-{1\over \alpha}(\omega^{2}-k^{2})^{4}.
\end{equation}
Hence one can see that the dispersion relation does not depend on
$\alpha$. The same holds for ghost modes, where the similar
condition reads
\begin{equation}
(\omega^{2}-k^{2})^{2}=0.
\end{equation}
Thus, we have obtained again the correct dispersion relation,
with the correct power of $(\omega^{2}-k^{2})$.

\end{document}